\documentclass[preprint,12pt]{aastex}

\shorttitle{zone region match}
\shortauthors{Fan et al.}
\usepackage{threeparttable}

\begin{document}

\title{Efficient Catalog Matching with Dropout Detection}

\author{Dongwei Fan\altaffilmark{1,2,3},
Tam\'as Budav\'ari\altaffilmark{2},
Alexander S. Szalay\altaffilmark{2},
Chenzhou Cui\altaffilmark{1}, 
and
Yongheng Zhao\altaffilmark{1}}

\altaffiltext{1}{National Astronomical Observatories, Chinese Academy of Sciences, 20A Datun Road, Chaoyang District, Beijing 100012, P.R. China}
\altaffiltext{2}{Department of Physics and Astronomy, The Johns Hopkins University, 3400 North Charles Street, Baltimore, MD 21218, USA}
\altaffiltext{3}{University of Chinese Academy of Sciences, 19A Yuquan Road, Shijingshan District, Beijing 100049, P.R. China}

\email{Email: dfan2@pha.jhu.edu}

\begin{abstract}
Not only source catalogs are extracted from astronomy observations. Their sky coverage is always carefully recorded and used in statistical analyses, such as correlation and luminosity function studies. Here we present a novel method for catalog matching, which inherently builds on the coverage information for better performance and completeness. A modified version of the Zones Algorithm is introduced for matching partially overlapping observations, where irrelevant parts of the data are excluded up front for efficiency. Our design enables searches to focus on specific areas on the sky to further speed up the process. Another important advantage of the new method over traditional techniques is its ability to quickly detect dropouts, i.e., the missing components that are in the observed regions of the celestial sphere but did not reach the detection limit in some observations. These often provide invaluable insight into the spectral energy distribution of the matched sources but rarely available in traditional associations.
\end{abstract}


\section{Introduction}
\label{sec:intro}
\noindent
With the technology improvements in modern telescopes over the last decade, astronomy now has large volumes of data that will soon reach the peta-byte regime. Novel data storage, automated processing and sophisticated publishing solutions are being
and will have to be developed to handle the challenges of ongoing and next-generation surveys, such as the Large Synoptic Survey Telescope \citep[LSST;][]{2006SPIE.6270E..24B}. Scientific progress is increasingly dependent upon our ability to analyze the immense amount of measurements to extract more information and ultimately develop an understanding of Nature's fundamental properties. 

Today many specialized instruments and facilities release invaluable data to the public on a regular basis. Multicolor and time-series studies rely on all of these observations.
One of the critical requirements of these analyses is automated tools that can meaningfully combine the data sets from several independent archives \citep{2001SPIE.4477...20M}. 
In particular, catalog matching has been the focus of several studies.
Some of these revisit the statistical aspects of finding the right discriminators for the associations \citep{2008ApJ...679..301B}, while others focus on the computational issues \citep{2005ApJ...622..759G, 2007cs........1171G, 2011ASPC..442...85P} to beat down the (naively) combinatorial nature of the problem.
These efforts are fundamentally influencing the way we handle observations and pave the road for next-generation analysis tools and services, such as those of the virtual observatories.

Folding in the sky coverage into catalog matching is key to efficient algorithms. Previous approaches either completely neglect to use this information or include them as an afterthought. By ignoring the footprints we eliminate our ability to constrain the brightness of non-detections that might appear in other observations. Without the coverage information, we simply cannot tell whether a missing match means a lower flux than the observational limit or the direction is simply outside the observed area. There are spatial query engines \citep[e.g.,][]{2001misk.conf..631K, 2004ASPC..314..289F} that can quickly find sources in a given region of the sky but they typically use a different indexing scheme to speed up the searches, and therefore require extra steps to re-organize the data set for the cross-matching.

Our approach is different. We create a procedure that incorporates spatial constraints directly into the crossmatching. It uses the same indexing scheme, hence performs better than previous techniques and can also deliver dropouts for further constraints on the spectral energy distribution of the sources.
In this paper, we introduce a solution that builds on the sky coverage information of the catalogs to exclude irrelevant parts of the data sets before the matching begins.
The method is introduced in Section~\ref{sec:method}. Section~\ref{sec:implement} describes the database implementation of the method, and extended to efficiently detect dropouts, Section~\ref{sec:perform} discusses the performance, and Section~\ref{sec:con} concludes our study.

\section{Matching with Spatial Constraints}
\label{sec:method}
\noindent
The Zones Algorithm is an efficient approach to crossmatching \citep{2007cs........1171G}.
The work presented here enhances its speed and functionality by adding geometrical constraints.
The original idea involves dividing the celestial sphere to (typically narrow) constant declination rings or bands called {\em{}zones}.
The detections are grouped into these pre-defined zones before matching.
Matching speedup comes from the fact that one can just consider sources in the same zone (as oppossed to the whole sky) and a small number of neighboring zones depending on the search radius.
When the search radius is shorter than the zone-height, one only has to consider two neighboring zones.
In addition sources are also sorted by the R.A. for further improvements (e.g., in processor cache locality).
This simple division has a great advantage over more complicated hierarchical schemes. Any given zone has only two direct neighbors unlike in the HTM or the HEALPix. It is also much easier to implement in any programming language and fits well with database engines.

For completeness, here we revisit the logical steps of an efficient database implementation. 
First we pre-calculate the zone identifier, the \textit{ZoneID} for each detection based on its declination $\delta$ as 
\begin{equation} \label{equzid}
{\rm{}\textit{ZoneID}} = \left\lfloor{}\frac{\delta+90^{\circ}}{h}\right\rfloor
\end{equation}
where $h$ is the zone-height in degrees and $\lfloor{}\cdot\rfloor$ represents the floor operation. Next sort the catalogs by \textit{ZoneID} and \textit{R.A.}, which typically consists of creating a table index.
The equatorial coordinates $(\alpha,\delta)$ of an object are converted to a unit vector with Cartesian coordinates $(x, y, z)$ following
\begin{eqnarray}
x &=& \cos\delta \, \cos\alpha \label{equx}\\
y &=& \cos\delta \, \sin\alpha \label{equy}\\
z &=& \sin\delta \label{equz}
\end{eqnarray}
Euclidean distance calculation with these coordinates is fast and angular separation thresholds map directly to distance limits.

This algorithm provides great improvement over naive implementations but it still searches through the entire datasets even if there is no overlap.
If one knew ahead of time what the overlapping parts of the sky are, the procedure could be further speed up.

\subsection{The Geometry of the Zones}
\label{subsec:zonegeom}
\begin{figure}
\epsscale{.30}
\plotone{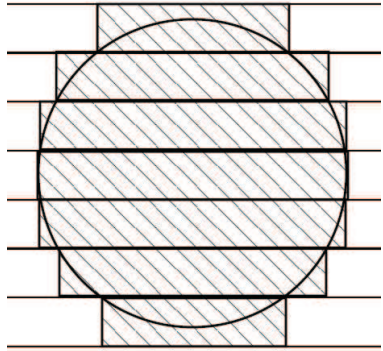}
\caption{Any shape on the celestial sphere can be approximated by the union of zone intervals. This cover will overshoot but only slightly considering that in practice we always use very thin zones.\label{f1}}
\end{figure}

\noindent
Survey geometry can be arbitrarily complex and their inclusion very involved. Software packages, such as the Spherical Toolkit \citep{2010PASP..122.1375B}, exist and can used to create, intersect and otherwise manipulate them. 
We can use these tools to approximate any geometry by the union of zone sections, as seen in Figure~\ref{f1}. 
The approximation will cover a somewhat larger area but will typically be adequate due to the narrow zones. The processing involves spherical geometry problems and is somewhat time consuming but has to be perform only once, if the same pre-defined zones are used for all catalogs.

\begin{figure}
\epsscale{.90}
\plotone{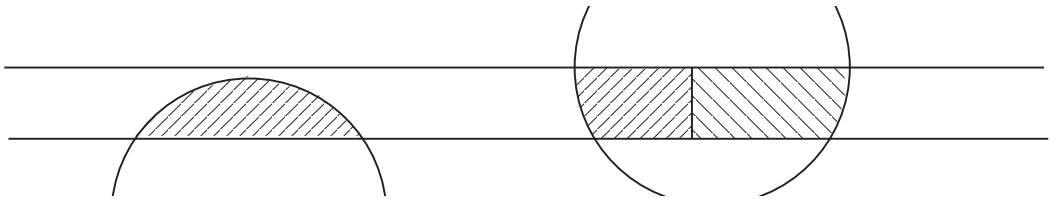}
\caption{Shapes described as generalized spherical polygons will usually consist of small-circle constraints. In these examples we show a few of the special cases that we have to handle. 
On the left we see that sometimes there are less than four vertices. The right hand side illustrates the wraparound problem. Circles across the meridian have to split and considered two separate shapes that yield disjoin zone intervals.\label{f4}}
\end{figure}

\begin{figure}
\epsscale{.45}
\plotone{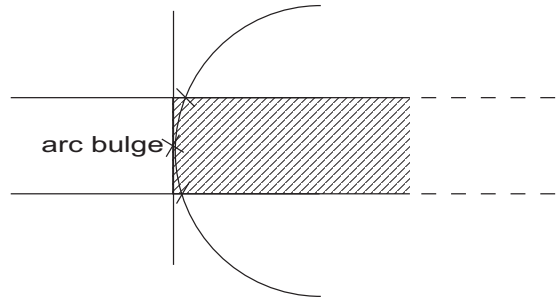}
\caption{This small-circle arc bulges into the zone beyond the vertices of the intersection. Special code is required to solve for these extrema and to extend the interval's endpoints accordingly.\label{f5}}
\end{figure}

\begin{figure}
\epsscale{.90}
\plotone{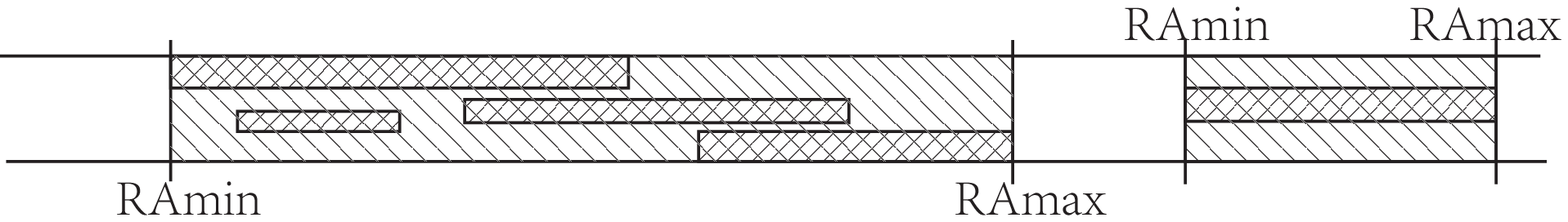}
\caption{The footprints may translate to overlapping zone intervals that can be simplified, which yields further performance enhancements. On the left we show such a complex situation in contrast to the usual trivial case on the right.\label{f6}}
\end{figure}

The implementation is straightforward but several noteworthy complications arise. Figure~\ref{f4} illustrates the possible intersections of a zone and different small circles. Due to the wraparound in R.A., a simply connected region could translate to more than one sections. Furthermore it is not enough to check the vertices of the intersections as small circles could protrude into the zone beyond the vertices as seen in Figure~\ref{f5}.
Complex geometry could produce fragmented intervals that overlap and could be simplified. Figure~\ref{f6} illustrates a scenario where such rectification is highly beneficial.
With these intervals in hand, the intersection of several footprints can quickly be determined with similar accuracy using the endpoints of the intervals.

\subsection{Matching with Sky Coverage}
\label{subsec:matching}
\noindent
The intersection area is approximated by a set of zone segments or intervals. Considering that the Zones Algorithm uses the same properties in the matching, namely the \textit{ZoneID} and (R.A., Dec.), one can introduce the new constraints directly into the matching code. These limits of the coordinates can be efficiently superimposed in databases that optimize the execution based on the indices and table statistics. The advantage over the traditional method is the explicit limits that trim the set of eligible combinations ahead of time.

Another great outcome of this approach is our new ability to find dropouts. When sources do not match anything in another catalog, one usually cannot tell whether there really is no counterpart for the given sensitivity of the specific instrument or the direction is simply outside the observational coverage. In our case the overlapping area is known and accurately represented, which can be used to decide. To distinguish we can use set operations readily available in database engines.

\section{Database Implementation}
\label{sec:implement}
\noindent
Today, the largest catalogs  typically reside in databases. These archives facilitate fast searches and provide different interfaces to the same data, which are frequently used by thousands of researchers every day. The SDSS Science Archive is arguably the archetypical example of such a database solution complete with a web-based user interface. In addition to form-based searches, power users can directly use the database engine by executing custom queries.

Our implementation of the new method was completely done in the Structure Query Language (SQL) and tested on the SDSS database. The geometrical computations were visualized and handled by the
the Footprint Service\footnote{\url{http://voservices.net/footprint/}}~\citep{2007ASPC..376..559B} and the Spherical Library\footnote{\url{http://voservices.net/spherical/}}~\citep{2010PASP..122.1375B}. 
The latter is a light-weight software package for high-performance calculations with arbitrary shapes on the celestial sphere and applicable on database system~\citep{2007cs........1163G}. In turn, the Footprint Service is bases on the Spherical Library. 
With these tools, the coverage of several telescopes and surveys have been created and recorded that include for example, SDSS, GALEX, and FIRST.

\subsection{Pre-processing in SQL}
\label{subsec:pre-processing}
\noindent
Before the crossmatch begins, several tasks are to be completed that include reorganizing and indexing the data. With a fix zone schema, the data sets can be prepared independently from each other. These pre-processing steps only have to be done once. We calculate the \texttt{ZoneID} and index the tables by that and the azimuth.
Another important part of the preparation is the processing of the sky coverage. The survey footprints are approximated by a union of zone intervals, whose derivation takes place in SQL followed by routines that simplify the description.
Each zone is bounded by two concentric small circles. The spherical shape that these describe are intersected with the footprint. Considering that the zones are typically very narrow, one could just look at the verticies but we handle the generic case. The issue is that small circles can bulge and protrude farther than the vertices.
Sometimes further simplifications are possible, as shown in Figure~\ref{f6}. In SQL we can combine together the intervals to avoid unnecessarily calculations.
In addition, we also have to deal with wraparound issues and split ranges across the meridian.

The results of the preprocessing are persisted in tables, which capture the approximate footprint and are ready to use in subsequent crossmatch queries. The schema saves the identifier and the interval for each zone: \texttt{ID, ZoneID, RAmin, RAmax}.

\subsection{Sky Coverage in SQL Matching}
\label{subsec:zonematch}
\noindent
Using the Zones Algorithm, one can reject the majority of the possible matches by considering only the relevant zones. Sources in two catalogs have to be considered if they are in the same zones. In addition to that, neighboring zones also can contribute matches, and zones farther away, if the search radius is longer than the zone height. These relations can be captured in a table that links the relevant zones called the \texttt{ZoneZone} table. 
Figure~\ref{codefig:zonealg} shows the original implementation of the matching using only the direct neighbors.
The variables \texttt{@theta} and \texttt{@dist2} represent the matching threshold in degrees and in Euclidean distance (squared), respectively.
We see that \texttt{Catalog1} and \texttt{Catalog2} are joined via the \texttt{ZoneZone} table to minimize the initial pairs. 
The \texttt{Alpha2} column in the \texttt{ZoneZone} table contains the amount of overshooting in  R.A. in the given zone such that the search can be completed without truncation.
As part of the JOIN clause, extra filtering on the angles provides fast rejection of sources that are too far and the Euclidean distance is only calculated in the WHERE clause for the best candidates.
The wraparound is also handled here by the last constraint in the JOIN clause. The source just before the meridian are also added on the other with negative values as part of the pre-processing. These conditions ensure uniqueness the pairs across the meridian.

We note that this query also includes a \textit{hint} for the SQL optimizer that is not obvious at first. In fact, the \texttt{LOOP JOIN} formally yields a suboptimal query plan, where the one catalogs is accessed in somewhat random order. The access pattern, however, is such that jumping happens in memory due to the ordering in R.A., cf. cache locality. This trick was invented by Jim Gray (private communication) and have been successfully used to create the largest crossmatches ever since.

In our new technique, the crossmatch query also includes the zone intervals, whose union represents the approximate shape of the common coverage of the observations. First this \texttt{ZoneIntersect} table is created, which involves testing the endpoints of the intervals. The process is straightforward but it is complicated by the wraparound problem. The SQL commands are relatively lengthy, hence they are omitted here.
The modified crossmatch query is shown in Figure~\ref{codefig:zonematch}.
The \texttt{ZoneIntersect} provides extra constraints on sources in \texttt{Catalog1}. It is also possible to subset the sources in the other catalog but the execution time does not improve as there are more limiting constraints already in place. Hints are also applied here to maximize the performance of the engine. These optimization tricks might prove unnecessary in future editions of SQL Server.

\begin{figure}
\begin{center}\begin{minipage}{8cm}{\footnotesize
\begin{verbatim}
SELECT c1.ObjID AS ObjID1, c2.ObjID AS ObjID2, ...
FROM Catalog1 AS c1
    INNER LOOP JOIN ZoneZone AS zz 
        ON zz.ZoneID1 = c1.ZoneID 
    INNER LOOP JOIN Catalog2 AS c2 
        ON zz.ZoneID2 = c2.ZoneID
        AND c2.RA BETWEEN c1.RA - zz.Alpha2 
                      AND c1.RA + zz.Alpha2
        AND c2.Dec BETWEEN c1.Dec - @theta 
                       AND c1.Dec + @theta
        AND ( c1.RA >= 0 OR c2.RA >= 0 )
WHERE (c1.Cx-c2.Cx) * (c1.Cx-c2.Cx)
    + (c1.Cy-c2.Cy) * (c1.Cy-c2.Cy)
    + (c1.Cz-c2.Cz) * (c1.Cz-c2.Cz) < @dist2
\end{verbatim}
}\end{minipage}\end{center}
\caption{The original crossmatch SQL query using the Zones Algorithm has been used to create the largest crossmatches. Its power comes from its simplicity and the clever ordering of tables. The zones restrict the number of possibilities up front and the index coordinates further reduce the problem. The Euclidean distance is only calculated for the nearby sources to perform the final cut, see text for details.\label{codefig:zonealg}}
\end{figure}

\begin{figure}
\begin{center}\begin{minipage}{8cm}{\footnotesize\begin{verbatim}
SELECT c1.ObjID AS ObjID1, c2.ObjID AS ObjID2, ...
FROM ZoneIntersect AS i
    INNER JOIN Catalog1 AS c1
        ON c1.ZoneID = i.ZoneID AND 
           c1.RA BETWEEN i.RAmin AND i.RAmax
    INNER LOOP JOIN ZoneZone AS zz 
        ON zz.ZoneID1 = c1.ZoneID
    INNER LOOP JOIN Catalog2 AS c2 
        ON zz.ZoneID2 = c2.ZoneID
        AND c2.RA BETWEEN c1.RA - zz.Alpha2 
                      AND c1.RA + zz.Alpha2
        AND c2.Dec BETWEEN c1.Dec - @theta 
                       AND c1.Dec + @theta
        AND ( c1.RA >= 0 OR c2.RA >= 0 )
WHERE (c1.Cx-c2.Cx) * (c1.Cx-c2.Cx)
    + (c1.Cy-c2.Cy) * (c1.Cy-c2.Cy)
    + (c1.Cz-c2.Cz) * (c1.Cz-c2.Cz) < @dist2
\end{verbatim}
}\end{minipage}\end{center}
\caption{The modified crossmatch query uses ideas from the Zones Algorithm but introduces spatial constraints that are fine-tuned for the indexing scheme. The geometry of the sky coverage is approximated by zone segments that provide fast filtering of the input data, see text for details.\label{codefig:zonematch}}
\end{figure}

\subsection{Dropout Detection}
\label{subsec:dropout}
\noindent
Different catalogs contain different sources. The commonality of several data sets is a function of the complicated selection effects. While our primary focus in crossmatching is to find counterparts, we also would like to know about non-detections, where the source is fainter than the limiting magnitude.

When a particular source in one of the catalogs has no counter part in another, we have to decide whether the given part of the sky is part of that observational coverage. Figure~\ref{codefig:dropout} illustrates the SQL command to achieve this and does it fast.
Here we look for dropouts in \texttt{Catalog1}. First we select objects that are inside the overlapping area, as \texttt{Objects1}, then we use the SQL \texttt{EXCEPT} keyword to perform the set operation: \texttt{Objects1} -- \texttt{MatchedObjects}. 

\begin{figure}
\begin{center}\begin{minipage}{8cm}{\footnotesize\begin{verbatim}
SELECT c.ObjID
FROM ZoneIntersect AS o 
	JOIN Catalog1 AS c ON o.ZoneID = c.ZoneID
   		AND c.RA BETWEEN o.RAmin AND o.RAmax
EXCEPT 
SELECT ObjID1 FROM MatchedObjects
\end{verbatim}
}\end{minipage}\end{center}
\caption{To select the \texttt{Catalog1}'s objects which are inside the specified region, then remove the objects that could be matched by \texttt{Catalog2}. The rest objects are the dropout objects of \texttt{Catalog1}.}
\label{codefig:dropout}
\end{figure}

\section{Performance Analysis}
\label{sec:perform}
\noindent
We tested the new method on the SDSS and GALEX catalogs (namely DR6 and GR3 AIS).
These catalogs contain 230.4 and 54.9 million detections, respectively, and their overlap on the sky is 3689 sq.~degrees. The matching radius was set to 7 arcseconds.
In the rest of this article, we quote running times and performance counters that were obtained on a single dual Intel Xeon E5430 CPUs (running at 2.66GHz) with 24 GB of memory.

After preparation the crossmatching was performed in 6 minutes 48 seconds using the original algorithm. The new algorithm completes in 5 minutes and 25 seconds, which is a significant 20\% improvement due to the small difference in the geometry. 
The new strategy excels even better in situations when the overlap is smaller.
We divided the data sets to create different test scenarios.
Considering that our method relies on the celestial coordinates, we generate cases by cutting on both coordinates to see the effect on the different shapes.

\begin{figure}
\plotone{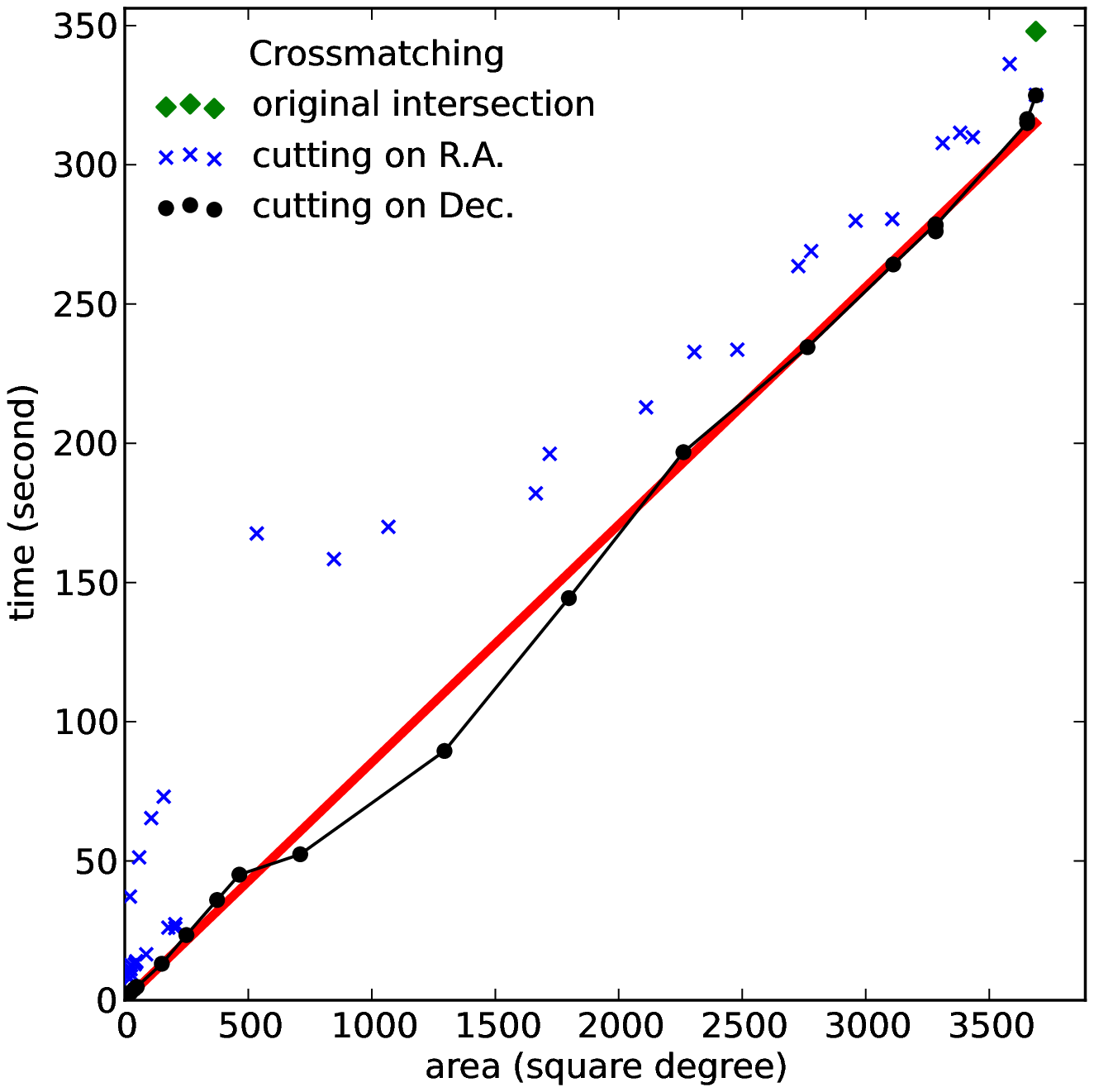}
\caption{The wall-clock time of the crossmatching scales linearly with the overlapping area. Cutting on the celestial coordinates corresponds to the worse and best case, hence we expect typical result somewhere between these in real-life situations. The red line plotted just to guide the eye.\label{f8}}
\end{figure}

Figure~\ref{f8} illustrates the elapsed time as a function of the overlapping area limited by R.A.\ and Dec.~ranges. The time scales linearly with the overlapping area. Black dots show the results for the subsets limited by Dec.; blue crosses represent tasks with limits in R.A. 
These two extremes are expected the bracket any real-life situation such as the actual intersection area of the two surveys shown with a green diamond. The red line through the origin is plotted to guide the eye.
Cutting on R.A.\ provides the lower bound of the performance. This is most noticeable at 500 sq.degrees. The regions in this case are very much elongated along Dec., hence they cut across many zones, which prevents the database from making the best use of the indexed tables.

\begin{figure}
\plotone{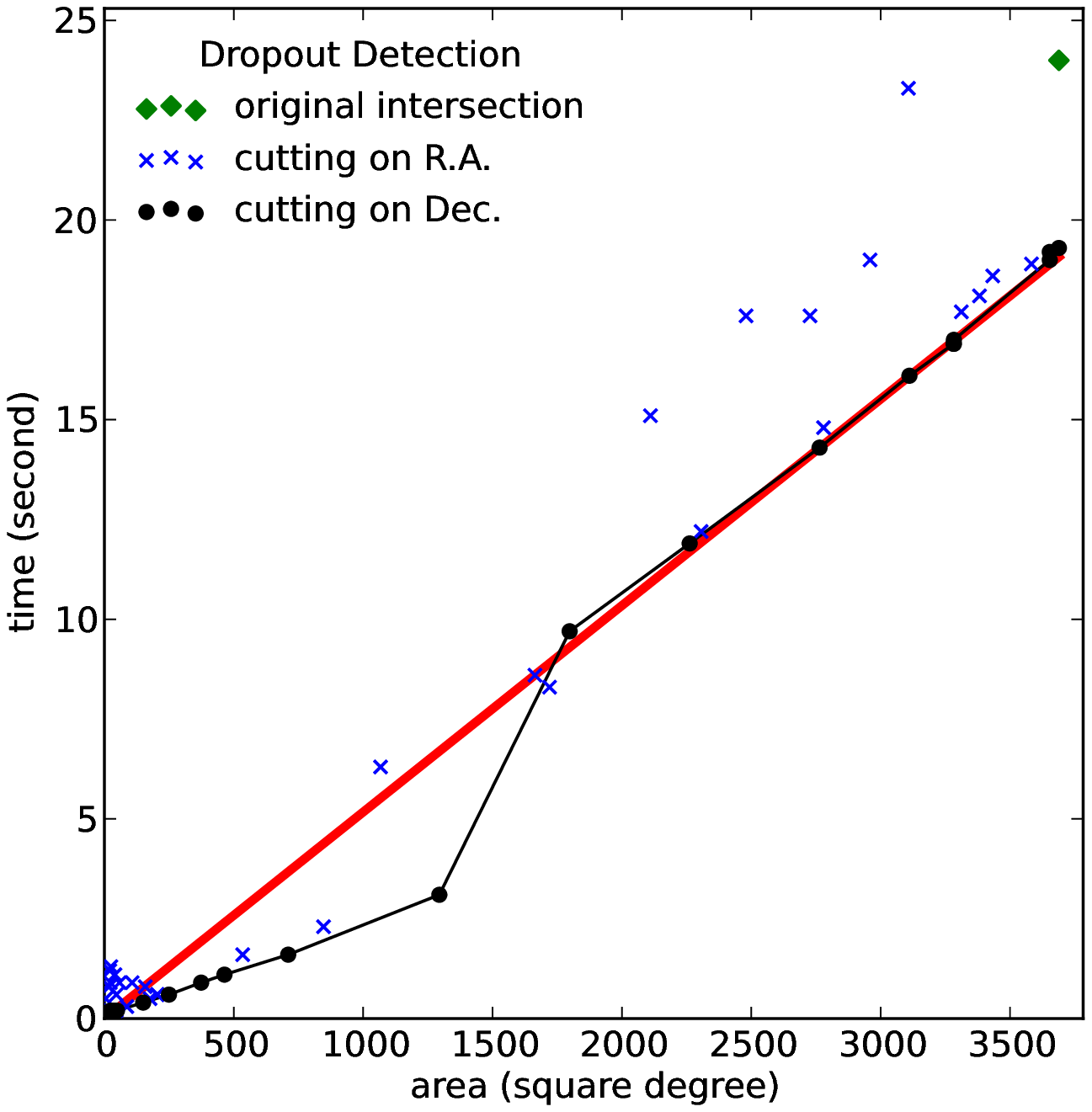}
\caption{Dropout detection can be efficiently performed after crossmatching the catalogs. The time is only a fraction of the matching and it also scales linearly with the overlapping area. The worse and best cases are plotted along with a trend line that guides the eye.
\label{f9}}
\end{figure}

After the crossmatching was completed, only a small fraction of that time is required to find the dropouts.
A total of 4.2 million GALEX dropouts, which are inside the SDSS coverage but do not have associated counterparts, were found in just 29 seconds.
Similarly to the previous test, we also consider the performance as a function of the overlapping area, and the results are are shown in Figure~\ref{f9}. The trend is again roughly linear with faster performance at areas below 1500 degrees. 
We made every attempt to clear the database's internal buffers using the \texttt{DBCC DROPCLEANBUFFERS} built-in procedure.
The sublinear performance, however, is likely to be a result of some other caching of the small amount of data involved, for example, in the hard drive. We did not try randomization to overcome this but expect the real-life performance to be somewhere on the red line.

\section{Conclusion}
\label{sec:con}
\noindent
We introduced a new algorithm for catalog matching that inherently uses the sky coverage information of all observations. The key feature of our approach is that it incorporates the coverage straight into the crossmatching and performs the filtering at runtime. It automatically restricts the datasets to the intersecting area to gain performance. User-defined spatial constraints can also be input for targeted matching.

The developed method builds on the Zones Algorithm. The footprints of the catalogs are approximated as R.A.\ intervals within the thin zones. The representation is not only highly accurate but also in line with the data organization required for efficient matching. Our implementation is in SQL, the Structure Query Language that is understood by most astronomy archives today including the SDSS, GALEX and HLA. Coding details are also provided along with measures of the performance scaling as a function of sky coverage.

The method also enables us to efficiently detect missing observations that were covered but did not appear over the detection threshold. It is a fast though preliminary step to identify dropouts. These dropouts are critical to constrain the spectral energy distribution of the celestial objects. Knowing that the flux is below a given limit can be very informative for a variety of studies ranging from photometric redshift estimation to galaxy evolution and cosmological analyses.

\section{Acknowledge}
\label{ack}
\noindent
This paper is funded by National Natural Science Foundation of China (10820002, 60920010, 90912005), Ministry of Science and Technology of China (BSDN2009-07), Beijing Municipal Science and technology Commission (2007A085).

\end{document}